\documentclass{emulateapj}
\usepackage{graphicx, color}
\usepackage{amssymb}

\newcommand{\pasa}{PASA}
\newcommand{\jcap}{J. Cosmology Astropart. Phys.}

\begin{document}
\shorttitle{Consistently exploiting magnified Type I\lowercase{a} SNe}
\shortauthors{Zitrin et al.}

\slugcomment{Submitted to the Astrophysical Journal}

\title{Consistent use of Type I\lowercase{a} supernovae highly magnified by galaxy clusters to constrain the cosmological parameters}

%\author{A. Zitrin\altaffilmark{1}}
%\author{J. Moustakas\altaffilmark{2}}
%\author{L. Bradley\altaffilmark{3}}
%\author{D. Coe\altaffilmark{3}}
%\author{L.A. Moustakas\altaffilmark{4}}
%\author{M. Postman\altaffilmark{3}}
%\author{X. Shu\altaffilmark{5}}

\author{Adi Zitrin\altaffilmark{1,2}, Matthias Redlich\altaffilmark{3}, Tom Broadhurst\altaffilmark{4,5}}

%\affil{
\altaffiltext{1}{Cahill Center for Astronomy and Astrophysics, California Institute of Technology, MS 249-17, Pasadena, CA 91125, USA; adizitrin@gmail.com}
\altaffiltext{2}{Hubble Fellow}
\altaffiltext{3}{Universit\"at Heidelberg, Zentrum f\"ur Astronomie, Institut f\"ur Theoretische Astrophysik, Philosophenweg 12, 69120 Heidelberg, Germany}
\altaffiltext{4}{Department of Theoretical Physics, University of Basque Country UPV/EHU, Bilbao, Spain}
\altaffiltext{5}{IKERBASQUE, Basque Foundation for Science, Bilbao, Spain}
%}

%\altaffiltext{*}{\textbf{The mass model and parameters are publicly available at: ftp://wise-ftp.tau.ac.il/pub/adiz/ElGordo}}}

\begin{abstract}
We discuss how Type Ia supernovae (SNe) strongly magnified by foreground galaxy clusters should be self-consistently treated when used in samples fitted for the cosmological parameters. While the cluster lens magnification of a SN can be well constrained from sets of multiple images of various background galaxies with measured redshifts, its value is typically dependent on the fiducial set of cosmological parameters used to construct the mass model to begin with. In such cases, one should not naively demagnify the observed SN luminosity by the model magnification into the expected Hubble diagram, which would then create a bias, but take into account the cosmological parameters a-priori chosen to construct the mass model. We quantify the effect and find that a systematic error of typically a few percent, up to a few-dozen percent, per magnified SN, may be propagated onto a cosmological parameter fit, unless the cosmology assumed for the mass model is taken into account (the bias can be even larger if the SN is lying very near the critical curves). We also simulate how such a bias propagates onto the cosmological parameter fit using the Union2.1 sample, supplemented with strongly magnified SNe. The resulting bias on the deduced cosmological parameters is generally at the few percent level, if only few biased SNe are included, and increasing with the number of lensed SNe and their redshift. Samples containing magnified Type Ia SNe, e.g. from ongoing cluster surveys, should readily account for this possible bias.
\end{abstract}

%\begin{keywords}
%\end{keywords}

%% text by Moustakas
%\def\imtxt#1{{\bf {\textcolor{red}{#1}}}}
%\def\imsout#1{{\bf {\textcolor{red}{\sout{#1}}}}}

\section{Introduction}\label{intro}

A Type Ia Supernova (SN) is an extremely luminous explosion of a star, typically a white dwarf in a binary system. Although there is still a debate regarding its exact progenitor mechanism \citep[e.g.][]{Maoz2012ProgenitorProblemReview}, an important property of a Type Ia SN is that its absolute peak luminosity is to a very good approximation well known (up to the Hubble constant, $M_{B}\approx-19.5$ mag, see \citealt{Riess1998,Hillebrandt2000SN} and references therein), thereby constituting a standardizable candle (e.g. taking into account the Luminosity-Decline Rate relation). In fact, it is thanks to this quality that we have learned greatly about the expansion of the Universe, particularly by comparing the standardized luminosities of many Type Ia SNe in different redshifts \citep[e.g.][]{Riess1998,Perlmutter1999}.

Clusters of galaxies act as strong gravitational lenses, distorting and magnifying background objects. When the surface mass density in the center of the cluster is high enough (higher than the critical density required for strong lensing; e.g. \citealt{NarayanBartelmann1996Lectures}), often \emph{multiple} images of the same background source are formed. Sets of multiple images in different redshifts are used therefore to constrain the underlying mass distribution and profile of the cluster's core \citep[e.g.][]{Broadhurst2005a,Smith2005,Limousin2008,Richard2010locuss20,Newman2013,Zitrin2013Gordo,Zitrin2013M0416}, dominated by an unseen dark matter (DM). Farther away from the center, where the surface density is lower, the gravitational potential of the cluster distorts and magnifies background objects (without forming multiple images of the lensed sources), and this weaker lensing effect can be used, statistically, to constrain the larger-scale mass distribution and profile of the cluster \citep[e.g.][]{Merten2009,Umetsu2010,Oguri201238clusters,Newman2013}. Lensing thus provides a unique way to map the DM in these massive objects.

Aside from mapping the unseen DM, lensing and especially the magnification by galaxy clusters has become of great interest, as it allows faint very-distant galaxies, which would otherwise be below the detection threshold, to be observed. Recent observations have made use of this magnification power to detect several compelling galaxy candidates at redshifts up to $z\sim10-11$ \citep{Coe2012highz,Zheng2012NaturZ,Bouwens2012highzInCLASH,Bradley2013}, and more are anticipated in the Frontier Fields program with the Hubble Space Telescope (HST)\footnotemark[6] \footnotetext[6]{http://www.stsci.edu/hst/campaigns/frontier-fields/}.

SNe which happen to explode in galaxies behind galaxy clusters, will therefore be magnified. In general, they are expected to appear in the same number density (or rate, see \citealt{Goobar2009LensedSN,Barbary2012magnifiedSN} and references therein) as in the field of a similar redshift, divided by the magnification factor which narrows the effective source-plane area, but supplemented by fainter or more distant SNe (for general discussion of the magnification bias see \citealt{Broadhurst1995MagBias,Mashian2013}), thus allowing to detect higher redshift SNe \citep[e.g.][]{BenitezRiess2002SN,Amanullah2011zp173SNmagnified,Barbary2012magnifiedSN,PanLoeb2013,Whalen2013HighzSNe}. As the current cosmological parameters are derived from a Hubble diagram of SNe up to $z\sim2$, higher redshift Type Ia SNe should tighten the constraints on the cosmological parameters.

Following \citet{Refsdal1964MNRAS}, many other works and dedicated surveys  (e.g. \citealt{KollattBartelmann1998lensedSN,Holz2001lensedSN,Goobar2002SNlensed,Goobar2009LensedSN,Oguri2003lensedSNandexpectation,Oguri2003lensedSNandH,Dawson2009lensedSN,Suzuki2012magnifiedSN,Riehm2011,Quimby2014arXivSciencemagnifiedSN} ; and references therein), have dealt with the possibility of observing a multiply-imaged SN, and the possibility of making use of measured time delays between the different multiple images to recover the Hubble constant, or other cosmological parameters. This is particularly appropriate for galaxy-scale lenses, where the time delay is observationally reasonable. In fact, time delays have been used in several studies to constrain the Hubble constant, making use typically of quasars multiply imaged by field galaxies (e.g. \citealt{Suyu2010MeasuredH,Suyu2013measured}, see also \citealt{Oguri2007timedelayssummary,Treu2013} and references therein).

Some of the works mentioned above have also referred to, or uncovered, a single image (i.e. not multiply lensed) of a SN magnified by a cluster, but only in the context of adding a constraint to the mass model through the local independent estimate of the magnification in the case of a Type Ia \citep[e.g.][]{Riehm2011,Nordin2013SN3}, or vice versa, using the magnification by the lens model to recover the SN demagnified luminosity (assuming a priori a set of cosmological parameters, e.g. \citealt{Patel2013SN3}, see also \citealt{Suzuki2012magnifiedSN,Amanullah2011zp173SNmagnified}).

%\citet{Oguri2003lensedSNandH} showed that the standardized-candle nature of Type Ia SNe, effectively breaks the degeneracy between the lensing potential (or mass profile) and the Hubble constant, a property which is needed especially for time-delays in galaxy-scale lenses where not many sources in different redshifts are lensed by a single lens (e.g. \citealt{Oguri2003lensedSNandH}, see also \citealt{Holz2001lensedSN,Wucknitz2002degenrqacies,SchneiderSluse2013}).

If highly magnified SNe were then to be used as part of samples fitted for the cosmological parameters, one should not naively demagnify the lensed SN luminosity by the magnification factor given by the mass model, but take into account the cosmological parameters that were used to construct it. The idea is quite simple in essence: one usually makes use of the fact that the mass-sheet and profile degeneracies are effectively already broken by various sets of multiple images typically uncovered in e.g. deep HST observations of cluster fields \citep[e.g.][as few examples; see also references therein]{Broadhurst2005a,Smith2005,Limousin2010M1423,Richard2010A370,Zitrin2012CLASH1206}, to construct a magnification map, determining the magnification of background objects such as lensed SNe in our case \citep[e.g.][]{Amanullah2011zp173SNmagnified}. However, since there is a degeneracy between the cosmological parameters and the resulting mass-model profile, which is typically left free to be fit by the data, this magnification is dependent on the cosmological parameters initially used to constrain the mass model. In such cases, to avoid circularity, one could use a simple analytic correction as the one we propose here as one example, simultaneously while fitting for the cosmological parameters, in order to disentangle the magnification value from the pre-assumed cosmology. Alternatively, one could simply take into account the possible systematic uncertainty induced by ignoring this effect; an uncertainty which we make an effort to quantify.

Because the resulting mass profile is dependent on the assumed cosmological parameters, several works \citep[e.g.][]{Jullo2010,Lefor2013} have shown that parametric strong-lensing (SL) or mass modeling techniques can be quite sensitive to the lensing distance of multiply-imaged sources, thus allowing to actually constrain the cosmological parameters. On the other hand, other works have shown that this dependence is rather weak \citep[e.g.][]{Zieser2012}, and more recent works have claimed to break or bypass the degeneracy between the profile and cosmological parameters, constraining them in a free-form modeling with minor assumptions on the mass profile shape; see \citet{Lubini2013Freeform,SerenoParaficz2013Freeform}. On a different front, \citet{Jonsson2010MNRASsigmafromlensedSNJ}, for example, exploited a large sample of Type Ia SNe magnified by foreground galaxies, to place constraints on the halos of the lensing galaxies, while fixing the cosmology and the mass profile shape (see also \citealt{Karpenka2013}).

Here, given recent and ongoing cluster surveys designed to detect strongly-magnified (and not necessarily multiply-imaged) SNe, which due to their magnification are also likely to expand the known Type Ia SNe redshift range \citep[see also][]{BenitezRiess2002SN,Amanullah2011zp173SNmagnified,Salzano2013SNe}, we highlight, as mentioned, how these strongly-magnified SNe should be properly treated when eventually used in samples fitted for the cosmological parameters (for example, some SNe more weakly magnified by galaxy clusters were used for that purpose as part of the Union2.1 sample, see \citealt{Suzuki2012magnifiedSN}), so that no bias is propagated from the cosmology assumed a-priori when constructing the lens model.

Many works have shown that a similar magnification or cosmology correction is also needed, statistically, when treating large samples of weakly magnified field SNe (e.g. \citealt{Linder+1988, Wambsganss+1997, HolzWald1998, SchmidtB+1998, Bergstrom2000SN, HolzLinder2005, Linder+1988, Sasaki1987,Martel2008, Amendola2013SNe,Marra2013SNe,Quartin2013SNesigma8}), suggesting how one should correct for the global magnification effect on their PDF in order to avoid a bias on the observed SN distance-redshift relation and the inferred cosmological parameters (see also \citealt{SmithM2013SNedistancesWL, Amanullah2003LensCorr}).

We aim to show that also small numbers of strongly magnified Type Ia SNe, and especially since these are expected to be observed to higher redshifts, can be useful as part of a sample fitted for the cosmological parameters, \emph{independently} of the cosmology assumed for the lens model (but still depending on the mass model parametrization). The presented methodology, although basic, evaded to our knowledge any discussion in previous works in this context (but some works have indeed properly quoted the cosmological parameters used to derive the magnification of magnified SNe, e.g. \citealt{BenitezRiess2002SN}). For our purpose, for simplicity, and since galaxy clusters are known to locally follow such mass profile forms (e.g. NFW \citealt{Navarro1996}; see also \citealt{Broadhurst2005a,Zitrin2009_cl0024,Umetsu2012}), we shall examine a simplified case by approximating the cluster mass profile in the strong-lensing regime, which is the area of interest in this work, with a power-law. This could be then generalized in future works.

This brief work is organized as follows: in \S 2 we show the dependence of the fitted mass profile on the cosmological parameters, and present a simplified method to correct the cosmology-dependent magnification of Type Ia SNe. In \S 3 we discuss the magnitude of the effect or bias in question, both on individual SNe and when propagated onto the Union2.1 sample supplemented with mock lensed SNe, and conclude the work, then summarized in \S \ref{summary}.

\section{Methodology}\label{Method}

The reduced deflection angle due to a given mass distribution, in the thin lens approximation, at a position $\vec\theta$ is given by:
\begin{equation}
\label{deflectiongen}
 \vec\alpha(\vec\theta)= \frac{4G}{c^2}\frac{d_{ls}d_{l}}{d_{s}} \int\frac{(\vec\theta-\vec\theta')\Sigma(\vec\theta')}{|\vec\theta-\vec\theta'|^{2}}d^{2}\vec\theta',
\end{equation}

where $d_{l}$, $d_{s}$, and $d_{ls}$ are the cosmology-dependent lens, source, and lens-to-source angular diameter distances, respectively, and $\Sigma$ is the projected surface mass density distribution.

Eq. \ref{deflectiongen} manifests the degeneracy between the lensing distance $\frac{d_{ls}d_{l}}{d_{s}}$, and hence the cosmological parameters, and the mass distribution. Correspondingly, lens modeling in complex systems such as galaxy clusters comprising various sets of multiple images, typically requires one to assume a set of cosmological parameters, while leaving the mass-density profile free to be fitted for. The magnification estimate for positions and background-source redshifts different from the lensing observables used as constraints, is thus cosmology dependent.

Imagine a background SN is observed at an angular distance $\theta_{SN}$ from the center of a (hereafter for simplicity, spherically symmetric) massive cluster. We now show explicitly how the mass profile, and thus the magnification, depend on the assumed cosmology; a dependence which can in turn be used to self-consistently rescale the magnification with the cosmological parameters. In what follows, two Einstein radii and enclosed masses ($M(<\theta_{e,i})$ and $M(<\theta_{e,j})$, $i \neq j$), are sufficient to show the said dependence. %The Einstein radii are of course cosmology independent and are observables, or deduced directly form them, and the masses are obtained by assigning a cosmology.

%Note also that the presented procedure is limited to the SL regime, so that $\theta_{SN}$ by definition cannot significantly exceed $\sim2\theta_{e,max}$, i.e. the region where multiple images may be observed (extrapolation of the forthcoming relations outwards to the weak lensing regime are possible, but may introduce a stronger model-dependent bias and are therefore riskier; we leave proper examination of the weaker-lensing regime to future studies).

\subsection{Example lens: a power-law}
A surface density power-law profile can be written as $\Sigma(r)=\Sigma_{0}(\frac{r}{r_{0}})^{-q}$, where $r$ is the physical distance from the center, and $r_{0}$ an arbitrary, normalization scale radius. The mass enclosed within an angular distance $\theta$ is obtained by integration of the latter density profile, while remembering that $r=d_{l}\theta$ and $r_{0}=d_{l}\theta_{0}$,  to obtain $M(<\theta)=\frac{2\pi\Sigma_{0}(d_{l}\theta_{0})^{2}}{2-q}(\frac{\theta}{\theta_{0}})^{2-q}$. The general deflection angle is given by:

\begin{equation}
\label{deflection}
 \alpha(\theta)= \frac{4GM(<\theta)}{c^2\theta}\frac{d_{ls}}{d_{s}d_{l}},
\end{equation}

or more explicitly by inserting $M(<\theta)$ from above:
\begin{equation}
\label{deflectiona}
 \alpha(\theta)=\frac{8\pi G\Sigma_{0}\theta_{0}}{(2-q)c^2} \frac{d_{l}d_{ls}}{d_{s}}(\frac{\theta}{\theta_{0}})^{1-q}.
\end{equation}

For a circularly symmetric lens, the dimensionless surface mass density, shear, and magnification at each position $\theta$ are generally given by, respectively:
\begin{equation}
\label{kappaeq}
\kappa_{(\theta)}=\frac{1}{2}\left(\frac{\alpha(\theta)}{\theta}+\frac{d\alpha(\theta)}{d\theta}\right),
\end{equation}
\begin{equation}
\label{gammaeq}
\gamma_{(\theta)}=\frac{1}{2}\left(\frac{\alpha(\theta)}{\theta}-\frac{d\alpha(\theta)}{d\theta}\right),
\end{equation}
\begin{equation}
\label{mueq}
\mu_{(\theta)}^{-1}=(1-\kappa)^{2}-\gamma^{2} ,
\end{equation}
where for a power-law surface density as above, the term $\frac{d\alpha(\theta)}{d\theta}$ simply equals $\frac{\alpha(\theta)}{\theta}(1-q)$. Plugging in now eqs. \ref{kappaeq} and \ref{gammaeq} into eq. \ref{mueq}, one obtains:
\begin{equation}
\label{mueq2}
\mu_{(\theta)}^{-1}=1+(q-2)\frac{\alpha(\theta)}{\theta}+(1-q)(\frac{\alpha(\theta)}{\theta})^{2} ,
\end{equation}
and $\alpha(\theta)$ is given in eq. \ref{deflectiona}.\\

The Einstein radius for a given multiply-imaged galaxy, is given (in the spherically symmetric case for example) by:
\begin{equation}
\label{einsteinrad}
\theta_{e}=\left(\frac{4GM(<\theta_{e})}{c^{2}}\frac{d_{ls}}{d_{s}d_{l}}\right)^{1/2} ,
\end{equation}

where more generally for a non-spherical case, the effective Einstein radius can be defined either as the radius within which $\langle\kappa\rangle=1$, or preferably (e.g. \citealt{Bartelmann1995arcs}), simply as the effective radius of the area enclosed within the critical curves for the redshift of the multiply-imaged galaxy. The Einstein radii are in any case observables, or deduced directly from them, and thus independent of the assumed cosmology, while the enclosed mass is cosmology dependent.

Having two measurements of the enclosed mass at e.g. $M(<\theta_{e,i})$, and $M(<\theta_{e,j})$, say, from the lens model constructed using various sets of multiple images and assuming a certain cosmology, a power-law mass profile could be readily fitted by:
\begin{equation}
q=2-\frac{\log\left(\frac{M(<\theta_{e,i})}{M(<\theta_{e,j})}\right)}{\log(\frac{\theta_{e,i}}{\theta_{e,j}})} ,
\end{equation}

and

\begin{equation}
\Sigma_{0}=\frac{2-q}{2\pi(d_{l}\theta_{0})^{2}} M(<\theta_{e,i}) (\frac{\theta_{e,i}}{\theta_{0}})^{q-2} .
\end{equation}

\subsection{Correcting for the assumed cosmology}

Recall from eq. \ref{einsteinrad} that for a given Einstein radius $\theta_{e}$, the mass enclosed inside $\theta_{e}$ is linear in the term defined hereafter as $D=\frac{d_{\rm l}d_{\rm s}}{d_{\rm ls}}$, so that $M(<\theta_{e})\propto D$, or explicitly:
\begin{equation}
M(<\theta_{e,i})=\frac{\theta_{e,i}^{2}c^{2}}{4G}D_{(z_{i})},
\end{equation}

where $z_{i}$ is the redshift of the lensed source galaxy whose Einstein angle is $\theta_{e,i}$, and $D$ depends on the respective lens- and lens-to-source distances.

The cosmological parameters affect the physical units of the mass model through $D$, via the angular diameter distances:
\begin{equation}
d_{(z_{a},z_{b})}^{A} = \frac{c/H_{0}}{1+z_{b}} \int_{z_{a}}^{z_{b}} dz \left(\Omega_{m}^{(0)}\cdot(1+z)^{3} + \Omega_{\Lambda}(z)\right)^{-1/2},
\end{equation}
for a flat two-component universe as an example.

Therefore, the \emph{modified} enclosed mass, $M'$, meaning the mass given a modified set of cosmological parameters embedded in $D'$, is then given by:
\begin{equation}
M(<\theta_{e,i})'=\frac{\theta_{e,i}^{2}c^{2}}{4G}D_{(z_{i})}',
\end{equation}

or,

\begin{equation}
M(<\theta_{e,i})'=M(<\theta_{e,i})\frac{D_{(z_{i})}'}{D_{(z_{i})}} ,
\end{equation}

where $M(<\theta_{e,i})$ is the Einstein mass of the $i$th system, with the set of cosmological parameters used to constrain the mass model to begin with.

Making use of the above, the \emph{modified} power-law mass profile, i.e. \emph{as if} the mass model were constructed with any other given set of cosmological parameters embedded in the term $D'$, can be readily calculated as:
\begin{equation}
\label{q_eq}
q'=2-\frac{\log\left(\frac{M(<\theta_{e,i})\frac{D_{(z_{i})}'}{D_{(z_{i})}}}{M(<\theta_{e,j})\frac{D_{(z_{j})}'}{D_{(z_{j})}}}\right)}{\log(\frac{\theta_{e,i}}{\theta_{e,j}})} ,
\end{equation}

and
\begin{equation}
\label{sigma_eq}
\Sigma_{0}'=\frac{2-q'}{2\pi(d_{l}\theta_{0})^{2}} M(<\theta_{e,i})\frac{D_{(z_{i})}'}{D_{(z_{i})}} (\frac{\theta_{e,i}}{\theta_{0}})^{q'-2}.
\end{equation}

From this, the ``new'', corrected magnification can be immediately calculated via eq. \ref{deflectiona}-\ref{mueq}, inputting $q'$ and $\Sigma'$ instead of $q$ and $\Sigma$, respectively.

This result is discussed further in \S 3.

\begin{figure*}
 \begin{center}
   \includegraphics[width=160mm,trim=5mm 5mm 5mm -12mm,clip]{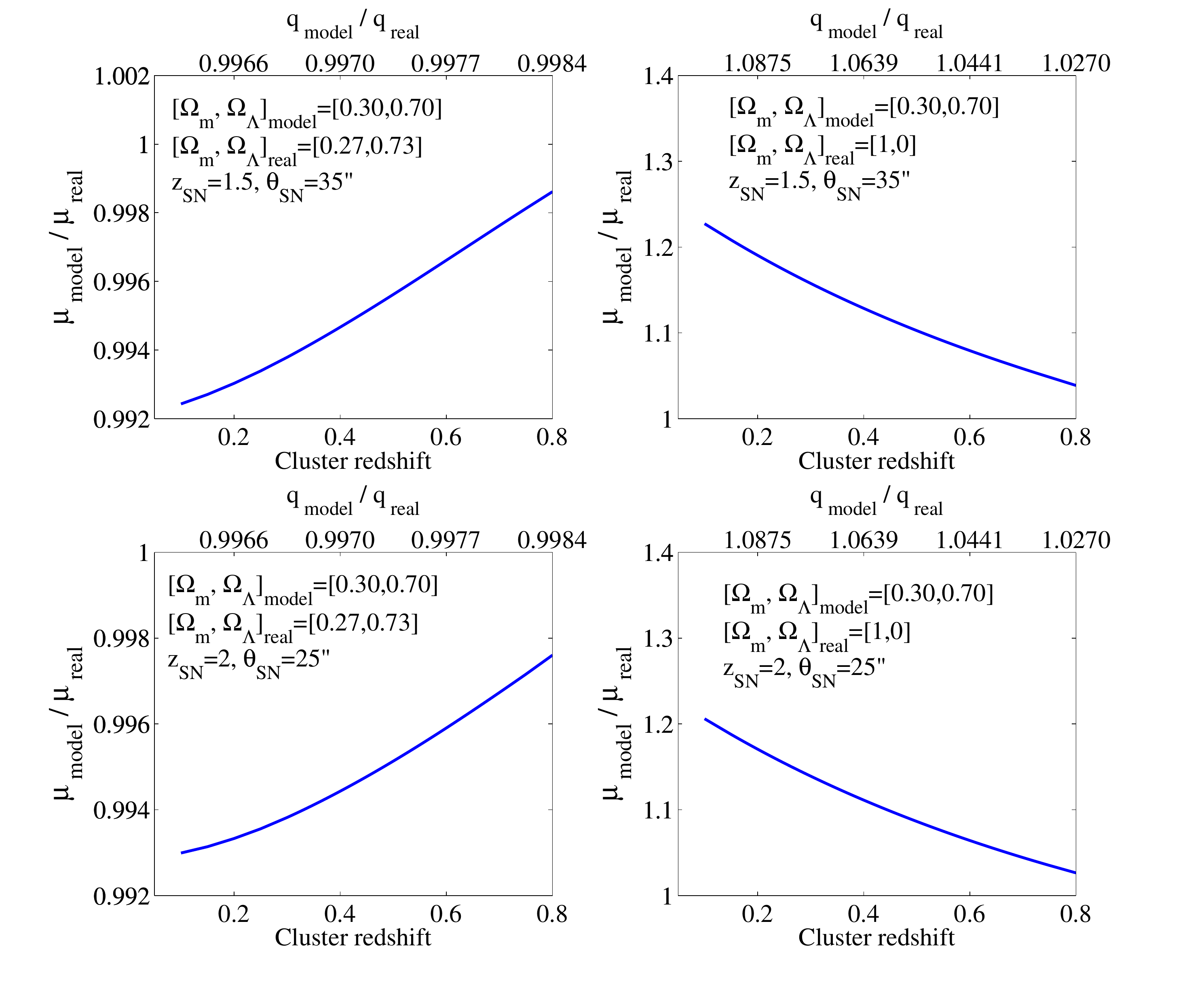}
 \end{center}
\caption{Bias created on the estimated luminosity of a lensed SN if the cosmology assumed for the lens model differs from the true underlying cosmology, as a function of lens redshift, for different input configurations. The different configurations are noted on each subfigure, such as the cosmologies used, the SN redshift ($z_{SN}$), and its distance from the center ($\theta_{SN}$). The upper $x$-axis shows the ratio of the power-law exponent given the ``modified'' cosmology and the power-law exponent given the ``true'' cosmology, for each configuration. In all cases we assume a circularly symmetric lens with two Einstein rings observed at $\theta_{e,1}=10\arcsec$ and $\theta_{e,2}=20\arcsec$, of sources at $z_{1}=1$ and $z_{2}=2$, respectively.  As can be seen, assuming for the lens modeling a cosmology which is only $\sim10\%$ different from the ``true'' underlying cosmology, results here in a minor $<1\%$ bias. However, more extreme differences between the assumed and probed cosmologies, can yield significant systematic errors of $\sim20\%$ on the demagnified SN luminosity, decreasing with lens redshift.}
\vspace{0.3cm}
\label{fig1}
\end{figure*}

\begin{figure*}
 \begin{center}
   \includegraphics[width=160mm,trim=5mm 5mm 5mm -12mm,clip]{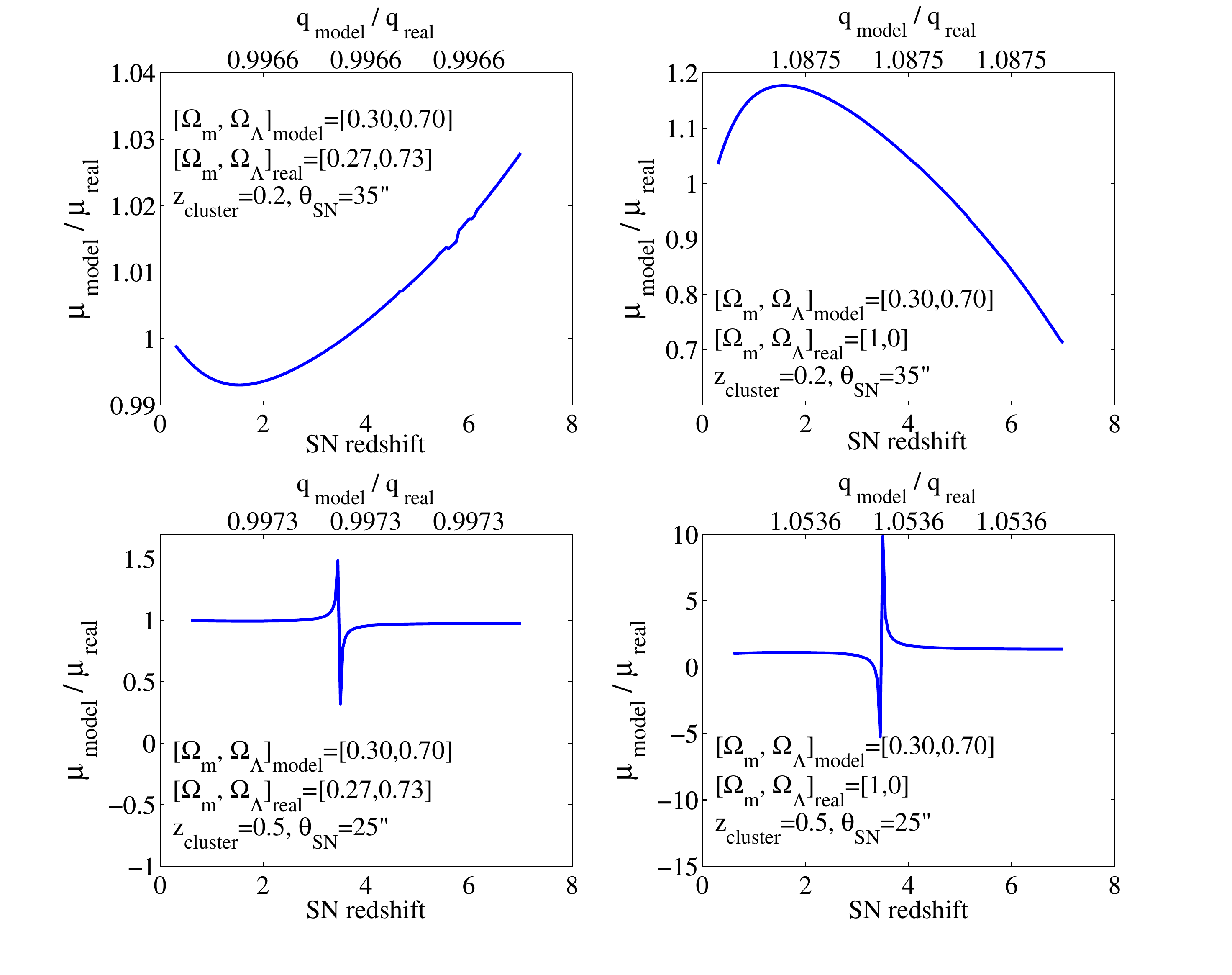}
 \end{center}
\caption{Same as Fig. \ref{fig1}, now showing the magnitude of the effect as a function of the magnified SN redshift, for various configurations. The \emph{top panel} shows a case where the SN is well outside the Einstein radius. As in Fig. \ref{fig1}, assuming for the lens modeling a cosmology which is only $\sim10\%$ different from the ``true'' underlying cosmology, results typically in a minor, few-percent bias. More extreme cosmology differences can yield significant systematic errors of about $\sim20\%$ in the cases probed here.  The \emph{bottom panel} case shows that if the SN is closer to the center, or near the narrow critical curves more explicitly, the effect can be much larger, reaching hundreds of percent. More importantly, the bias can reach up to $\sim50\%$ (depending on the cosmology difference) for higher redshift SNe, and especially if they lay within the critical curves for that redshift. Note that in this figure, the ratio of exponents of the ``modified'' and ``true'' cosmology mass models (top $x$-axis), is constant, because as expected the shape of the lens is not affected by the SN position or redshift.}
\vspace{0.3cm}
\label{fig2}
\end{figure*}

\begin{figure*}
 \begin{center}
   \includegraphics[width=160mm,trim=5mm 5mm 5mm -12mm,clip]{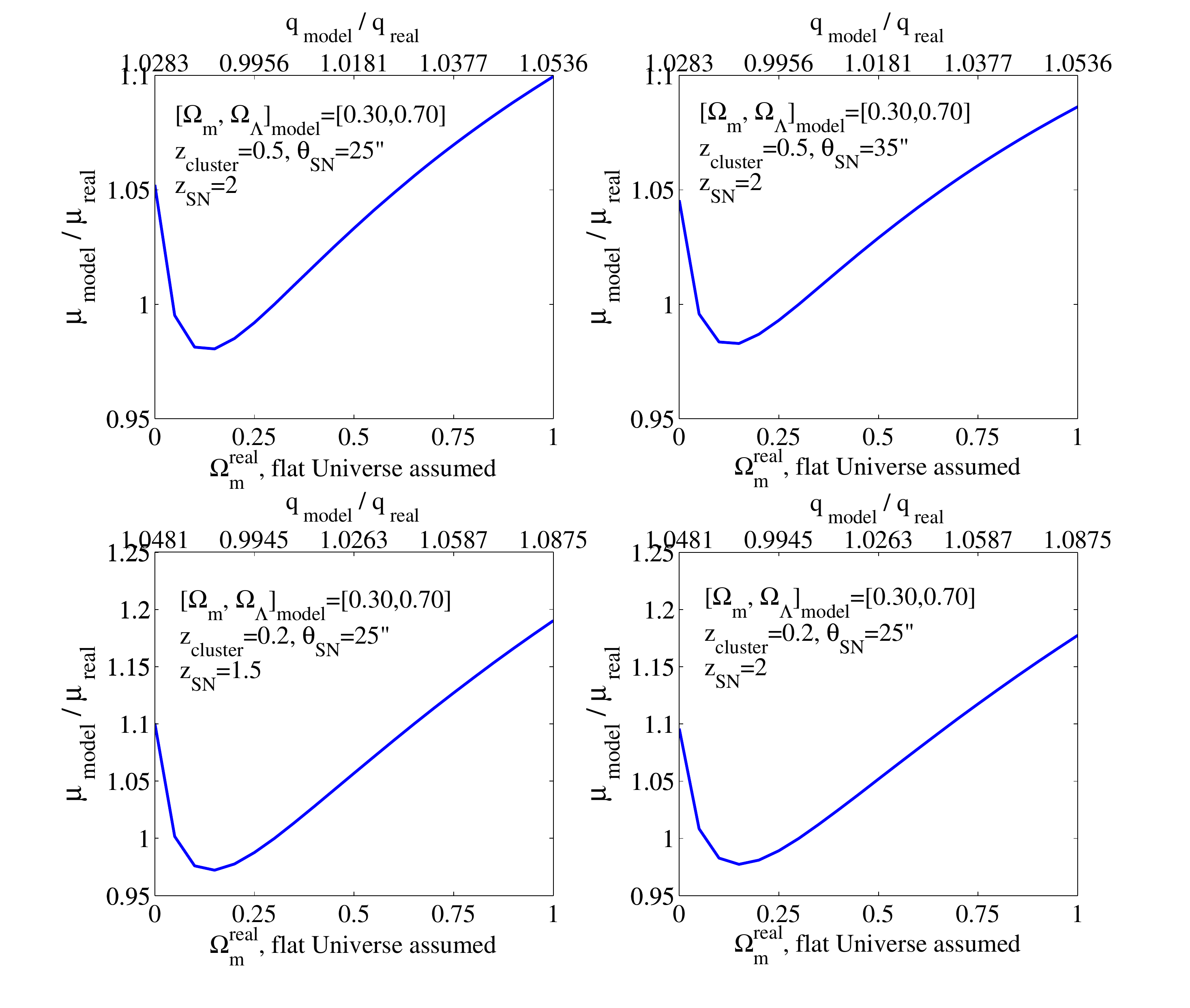}
 \end{center}
\caption{Same as Figs. \ref{fig1} $\&$ \ref{fig2}, but now fixing the lens configuration, and changing the ``true'' underlying (flat-universe) cosmology, whereas the cosmology used for constructing the model is unchanged. As expected, the bias vanishes ($\mu_{model}/\mu_{real}$ reaches unity) when the model and true cosmologies are similar, and is maximal when the cosmologies significantly differ. Note also that due to inherent degeneracy in the dependence of the magnification on the cosmological parameters, there can be, as seen, other cosmologies that yield occasionally similar magnification values (and thus zero bias, $\mu_{model}/\mu_{real}=1$).}
\vspace{0.8cm}
\label{fig3}
\end{figure*}

\section{Discussion and Conclusions}\label{Disc}

In \S \ref{Method}, we demonstrated the known degeneracy between a mass-model density profile, and the cosmological parameters. We have shown that by approximating the resulting mass profile with a known analytic form, the said degeneracy can be in turn used to self-consistently rescale the magnification estimate of a lensed SN, with the cosmological parameters. The approximation we showed is useful since it does not require remaking the usually-complex lens model for each probed set of cosmological parameters, which would be a hard and time consuming task (an order of hours on nowadays machinery, for each full-minimization iteration). Instead, one could use the above quick-to-calculate relation to readily obtain the SN magnification as a function of cosmology, given the initial mass model and the assumed fiducial set of cosmological parameters.

However, since the suggested correction is itself model dependent, it may be instead useful to simply account for the systematic uncertainty entailed by ignoring the cosmology assumed for the lens model. To estimate the magnitude of this bias, so that instead of using the above approximation, lensed SNe could be fitted for while not underestimating the uncertainties on their demagnified luminosities, one should examine the susceptibility of the magnification estimate to the cosmological parameters. This is shown in Figs. \ref{fig1}-\ref{fig3}, where we also give further explicit details. In Fig. \ref{fig1}, we plot the ratio between the magnification given a set of cosmological parameters used to construct the mass model, and the magnification obtained with the ``true'' cosmological parameters, for different configurations, as a function of cluster redshift. Fig. \ref{fig2} shows the same effect as in Fig. \ref{fig1}, now as a function of SN redshift, and Fig. \ref{fig3}, as a function of the difference between the underlying cosmology, and that assumed for the mass model. The magnitude of the bias clearly changes as a function of the observables (e.g. Einstein radii and source redshifts, SN position), cluster and SN redshift, and the difference between the cosmology assumed for the mass model and the ``true'' one. As seen, the magnitude of the bias created per lensed SN is typically of an order of a few percent, especially if the SN is observed at a larger angle, far from enough the Einstein ring, although some configurations can yield a bias of up to a few-dozen percent or higher, especially for lower-$z$ clusters, or if the SN is close to the center (or to the critical curves). This shows that the effect in question can, in principle,  be significant.

If the mass model was, as is often the case, constructed with cosmological parameters $\sim10\%$ away from the ``true'' parameters, the effect is typically less than $\sim1\%$ and thus rendered negligible. If the difference between the assumed cosmology and the true one, is higher, the bias can as significant as $\sim20\%$, per SN.

In that aspect, for comparison, we also mention that typical modeling errors of current high-end lens models (for fixed cosmology) are of order $\sim15-20\%$ on the magnification in most of the region of interest (i.e. not too close to the critical curves), and systematic errors between different parameterizations are typically of the same order. The bias we discuss in this work, as mentioned, is in most cases smaller, but still significant even in light of the non-negligible errors on the deduced magnification when a fixed cosmology was assumed. In the era of precision cosmology and with the expected numbers of SNe into higher redshifts, one should use these corrections not to create (even a small) bias, or alternatively, take into account the estimated systematic uncertainties. We also note that another alternative, would be making the mass model while allowing the cosmological parameters to be free, thus marginalizing over their effect on the magnification which would be reflected in the quoted errors.

To test how the cosmology affects real-cluster lens models, we chose one CLASH cluster with 2 spectroscopically measured multiply-imaged galaxies, at $z\simeq1.5$ and $z\simeq3$ (all CLASH mass models will be soon published in Zitrin et al., 2014 in prep, including the multiple images and exact redshifts). We then constructed two SL models, using the lens model code described in \citet{Zitrin2013Gordo,Zitrin2013M0416} which include realistic representations for both the cluster lens galaxies and the DM. For our purpose, the first model here is constructed using [$\Omega_{m}=0.3$, $\Omega_{\Lambda}=0.7$], and the second is constructed using a very distinct flat-universe cosmology, [$\Omega_{m}=1$, $\Omega_{\Lambda}=0$]. The different cosmologies, in practice, translate to a $\sim15\%$ difference in the effective, relative lensing distances. We then examined the resulting magnification maps in the 2x2 arcmin central FOV around the BCG. We find that with respect to the [$\Omega_{m}=0.3$, $\Omega_{\Lambda}=0.7$] model, the magnifications of the second, [$\Omega_{m}=1$, $\Omega_{\Lambda}=0$] model, deviate by 17.3\% on average throughout this FOV, and by a median of 1.9\%. These values increase as the FOV shrinks towards the SL regime, and the median reaches $\sim8\%$ in the central 1x1 arcmin field, which corresponds roughly to the SL regime of this test cluster. We take this median value as the more representative one (the mean here is much higher than the median because of the diverging critical curves), and conclude that in the SL regime, close to a (median value of) $\sim10\%$ bias in the magnification can be induced if the wrong cosmology is used. We note, however, that in this paper we focus on introducing the bias and assessing its order of magnitude, showing that in principle, it can be significant and should be taken into account. A more thorough estimate of the bias in real clusters, including also, for example, realistic SN distributions in redshift convolved with lensing models and a general cluster mass function, should be performed elsewhere.

We make an additional effort to examine how the possible bias on individual, lensed SNe, shown in Figs. \ref{fig1}-\ref{fig3}, propagates into the cosmological fit for a Union2.1-like sample. For that purpose we downloaded the Union2.1 sample\footnotemark[7] \footnotetext[7]{http://supernova.lbl.gov/Union/} and reran a cosmological fit to their data, starting by fitting the original data including the 580 SNe listed therein, and then supplementing it with increasing numbers of magnified SNe. For each minimization we run a simple a Monte-Carlo Markov Chain (MCMC) with Metropolis-Hastings algorithm, to obtain the best fit. Note that the minimization or best-fit criterion we use here is a simple $\chi^2$ defined as:

\begin{equation}
\chi^{2}=\sum_{SNe} \frac{\mu_{B}-\mu_{B,fit}}{\sigma_{err}^{2}}    ,
\label{chi2}
\end{equation}

where $\mu_{B}$ and $\mu_{B,fit}$ are\footnotemark[8] \footnotetext[8]{note that here $\mu_{B}$ are the distance moduli, while throughout, the lensing magnification is also marked as $\mu$ (i.e. without the capital ``B''), following traditional notation.} the observed distance modulus, and that predicted by the fit, respectively, and $\sigma_{err}$ is the error specified in the Union2.1 table available online. As a first test, the best-fit values for the original sample we obtain are:  $\Omega_{m}=0.2776^{+0.1421}_{-0.1032}$ and $w=-1.0005 ^{+0.1951}_{-0.4521}$ (1$\sigma$ errors). The best-fit values are in excellent agreement with those published in \citet{Suzuki2012magnifiedSN}, e.g., $w=-1.001^{+0.348}_{-0.398}$, albeit the errors are somewhat different, probably due to the difference in the $\chi^2$ definition and the inclusion of other systematics therein. Here however we only need to work in our self-consistent frame-of-reference to check the effect of including magnified SNe in the fit, on the resulting cosmological parameters. We note that the errors in the following scenarios we probe are all similar throughout, and we shall only focus on the difference between the best-fit values themselves, unless otherwise stated.

\begin{figure*}
      \includegraphics[width=174mm,trim=24mm 42mm 30mm 28mm,clip]{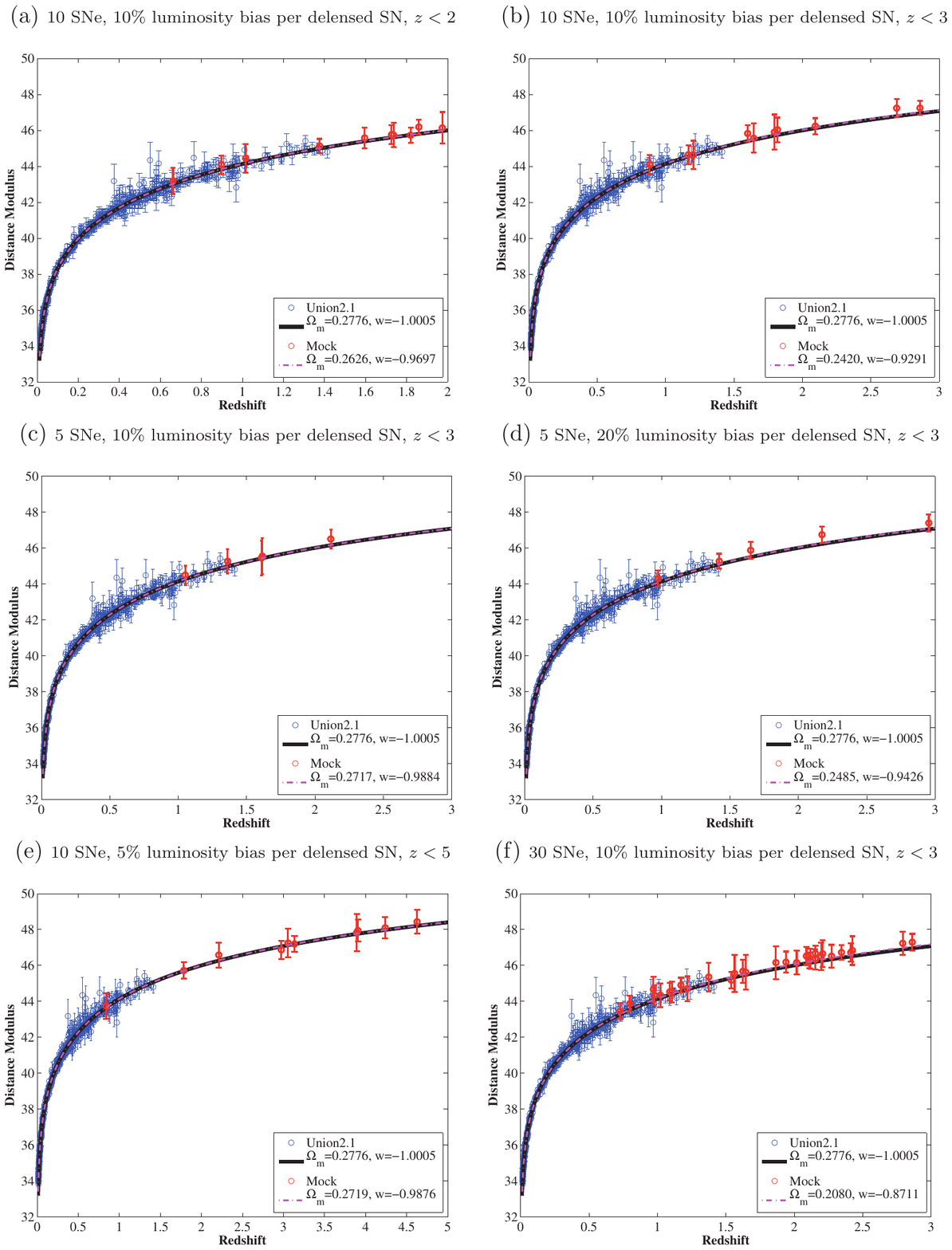} %trim [x down x up]
        \caption{Effect of the demagnified-luminosity bias discussed in this work, on the overall cosmological fit to the Union2.1 sample when supplemented with (de)lensed SNe. Figure shows different examples of mock, lensed Type Ia SNe (\emph{red error-bars}), on top of the Union2.1 sample (\emph{blue error bars}). The mock SNe imitate observed, magnified SNe, demagnified back to their unlensed luminosities with a magnification factor biased by the amount specified in each subfigure. This luminosity bias can be created if one neglects the cosmology assumed for the lens model (see Figs. \ref{fig1}-\ref{fig3}). The \emph{solid black lines} show our fit to the original Union2.1 sample, and the \emph{magenta dash-dotted lines} show the fit to the entire sample including the mock SNe. The corresponding best-fit values, assuming a flat universe and a fixed equation of state parameter, are shown in the \emph{Legends}, and demonstrate the overall bias created. We also specify for each subfigure some input restrictions used for generating the mock catalogs (see \S \ref{Disc} for more details).}\label{union}
\end{figure*}

After the initial fit we run to the original Union2.1 sample, we then plant SNe drawn from a uniform distribution between $z=0.5$ and up to either $z=1.5$, $z=2$, $z=3$, or $z=5$, for the different scenarios we consider (as mentioned, lensed SNe should in principle be observed to higher redshifts than field SNe). The SNe are planted following a distance modulus-redshift relation with the best-fit parameters from the initial fit to the full sample, and with a random Gaussian scatter of $\sigma=0.15$, and a random Gaussian error-scatter of $1\%+\sigma_{err}$, with $\sigma_{err}=abs(0.3)$, in their distance moduli. The luminosity bias propagated per demagnified SNe is taken as $5\%$, $10\%$, or $20\%$, for the different scenarios we examine here.  Examples of real+mock, distance-modulus vs. redshift relation are seen in Fig. \ref{union}.

The propagated bias on the overall fit turns out to be non-negligible, even with relatively small numbers of lensed SNe. Ten mock lensed SNe with a $10\%$ bias on the demagnified luminosity of each, for example, drawn from a distribution as described above up to $z=2$, create a shift (or bias) of $\simeq5\%$ and $\simeq3\%$ on the best fit $\Omega_{m}$ and $w$, respectively. Increasing the redshift upper limit to $z=3$ brings the overall bias to $\simeq13\%$ and $\simeq7\%$, respectively. When increasing the bias on each individual SNe demagnified luminosity to $20\%$, a $z<1.5$ sample yields a bias of $\simeq7\%$ and $\simeq3\%$ on the best fit $\Omega_{m}$ and $w$, respectively, and the $z<3$ sample yields a bias of $\simeq15\%$ and $\simeq8\%$ on the two parameters, respectively. Decreasing the number of lensed SNe to as few as five, or lowering the individual bias to $5\%$, reduces the overall bias by a few times, but tripling the number of lensed SNe to 30 up to $z=3$, can reach a large bias of $\simeq25\%$ and $\simeq13\%$ on the two parameters, respectively. Although in most probed cases the resulting bias is $<1\sigma$, some configurations yield biases that can be more significant, increasing with the individual bias on the demagnification factor, the number of lensed SNe, and their redshift.

As a final consistency check, we run two additional minimization chains while planting 20 higher-redshift SNe up to $z=5$ following our initial fit to the original Union2.1 sample. The first case includes unbiased SNe, and the second case includes SNe biased by $\sim10\%$ as above. The first chain results as expected, in cosmological parameters ($\Omega_{m}$ and $w$) identical to those obtained by the fit to the original Union2.1 sample, but with errors lower by $\sim15-20\%$, indicating, as expected, that including higher-redshift SNe improves the constraints on the cosmological parameters. In the second chain, we work on the sample containing the $\sim10\%$-biased mock SNe, but now take into account this additional systematic uncertainty in the fit, increasing the errors on the planted SNe, correspondingly, to include the $\sim10\%$ uncertainty originating from the bias. This, in order to examine, briefly, if including magnified (i.e. possibly biased) SNe in the fit is worthwhile. We obtain that the cosmological parameters are reproduced with a $>99.99\%$ accuracy, and the errors on them remain the same as for the original Union2.1 sample (but not smaller, despite including higher-redshift galaxies). This indicates that it is indeed worthwhile including magnified SNe in the fit, if the possible bias discussed in this work is accounted for as an additional error on their magnitude or distance modulus, so the resulting cosmological parameters will indeed remain unbiased.
\\

Since our goal here was simply to introduce the effect of the cosmological parameters assumed for constructing the mass model on the measured magnification, assess its order of magnitude, and show how it can be corrected for when fitting for the cosmological parameters, we brought one simple example using an idealized parametrization of a (circularly symmetric cluster) power-law mass profile, to rescale the magnification with cosmology. Clearly, this power-law approximation cannot describe the usually more-complex mass profile to a tee, and thus in practice, can create its own model-dependent bias (although we know from previous analyses the approximation is reasonable for the inner SL region, e.g. \citealt{Broadhurst2005a}; \citealt{Zitrin2009_cl0024}). Other, analytic or more flexible parameterizations, which may be better fitted per cluster, can be developed in future studies. To estimate the dependence of the magnification on the cosmological parameters more generally, these can include, for example, non-parametric (free-form mass profile) methods marginalizing over the cosmological parameters, or a Taylor-expansion of the magnification in the cosmological parameters.

The rate of SNe behind clusters, as mentioned, was examined before in various works \citep[e.g.][]{Sullivan2000,Barbary2012magnifiedSN,Goobar2009LensedSN,Riehm2011,PostmanCLASHoverview,LiHjorthJohan2012,Salzano2013SNe,Quartin2013SNesigma8,Graur2013arXiv1310.3495G}, and we gather that an order of magnitude of few Type Ia SNe within the HST's FOV are expected, per observed cluster with a typical depth of say, $\sim27$ AB spread over few years, with $\sim$weekly-to-monthly visits. However, as these are very crude numbers and depend exhaustively on the observational plan and lensing strength, we refer the reader to the works mentioned above for specific details. In our work here, we merely introduce and characterize the bias in question and do not attempt to assess its realistic distribution in Universe, following for example SN luminosity functions convolved with realistic mass models and a cluster mass function. We leave such estimates for future studies.

Lastly, one should also comment on the weak-lensing regime, in which the magnification is typically small, approaching $\simeq1$ in the outskirts of the cluster. Despite the smaller magnification, the significantly larger area covered by the weak-lensing regime (i.e. out to the virial radius and beyond) is advantageous, and large numbers of slightly-magnified SNe might be uncovered, to further reduce the statistical errors and form a useful representative sample, which could make use of similar corrections to those outlined here, albeit these are expected to be correspondingly smaller.

\section{Summary} \label{summary}
A cluster mass model constructed from various sets of multiple images can be used to estimate the magnification at the position where a highly magnified Type Ia SN is seen. If the demagnified SN brightness were then to be used as part of a sample fitted to constrain the cosmological parameters, to avoid a bias originating from the cosmology assumed for the lens model, the latter should be accounted for. We showed that in principle this can be done in a simple and elegant way, by approximating the resulting mass profile with a known analytic form.

More importantly, and especially since such a correction is by itself model dependent, we quantified the effect of ignoring the cosmology assumed for the lens model, on the magnification estimate of lensed SNe. We have found that a systematic error of typically a few percent, up to a few-dozen percent, per magnified SN, can be propagated onto a cosmological parameter fit, unless the cosmology assumed for the mass model is taken into account. In some specific cases, the bias can be even larger, for example if the SN is lying very near the critical curves.

We then simulated how such a bias, per SN, propagates onto the cosmological parameter fit using the Union2.1 sample, when supplemented with strongly magnified SNe. The resulting bias turns out to be non-negligible. We found that the bias on the deduced cosmological parameters is generally of the order of a few percent, if only few biased SNe are included, and increasing with the number of lensed SNe, their redshift, and the original bias from the lens model. Ultimately, we verified that the cosmological parameters are indeed accurately reproduced, if the bias on each magnified SNe is taken into account in the fit.

Several SNe magnified by galaxy clusters are already in hand (e.g. \citealt{Amanullah2011zp173SNmagnified}) or anticipated to be uncovered soon: several magnified Type Ia SNe were used in the Union2.1 compilation \citep{Suzuki2012magnifiedSN}; expected or recently found in the CLASH (\citealt{PostmanCLASHoverview,Salzano2013SNe,Patel2013SN3,Whalen2013HighzSNe}, see also \citealt{Graur2013arXiv1310.3495G}), and Frontier Fields programs; and many more are expected further ahead, for example with the James Webb Space Telescope \citep[JWST; e.g.][]{PanLoeb2013}. We conclude, especially given the leap over recent years in strong-lens modeling accuracy, that the effect calculated here should be readily taken into account with existing and soon to come data, to take proper advantage of magnified SNe when these are in turn used for constraining the cosmological parameters. In addition, the magnitude of  the effect investigated here can be useful for related purposes, such as for estimating the additional error on the derived magnification of lensed high-$z$ galaxies, originating from the choice of cosmological parameters used for the lens model.

\section*{acknowledgments}
We kindly thank the anonymous reviewer of this work for most valuable comments. AZ greatly thanks Miguel Quartin, Matthias Bartelmann, Richard Ellis, Steve Rodney, Massimo Meneghetti, Or Graur, Saurabh Jha, Brandon Patel, Keiichi Umetsu and Matt Schenker, for useful discussions and comments. Support for AZ is provided by NASA through Hubble Fellowship grant \#HST-HF-51334.01-A awarded by STScI. Part of this work was supported by contract research ``Internationale Spitzenforschung II/2-6'' of the Baden W\"urttemberg Stiftung.

%\bibliographystyle{apj}
%\bibliography{outDan2}

\end{document}